\documentclass[9pt,a4paper]{article}
	\usepackage{graphicx}
	\usepackage{color}
 	\usepackage{fancyhdr}
\usepackage[super,sort&compress,comma]{natbib} 
\usepackage[version=3]{mhchem}
\usepackage{balance}
\usepackage{times,mathptmx}
\usepackage{graphicx} 
\usepackage{lastpage}
\usepackage{float}
\usepackage[english]{babel}
\usepackage{array}
\usepackage{droidsans}
\usepackage{charter}
\usepackage[T1]{fontenc}
\usepackage[usenames,dvipsnames]{xcolor}
\usepackage{hyperref}
\usepackage[margin=0.5in]{geometry}
\usepackage{epstopdf, epsfig}
\usepackage{upgreek}
\usepackage{relsize}
\pdfoutput=1
\begin{document}

\LARGE{\textbf{Gas bubble evolution on microstructured silicon substrates $^\dag$}} \\
\vspace{1cm}

\large{Peter van der Linde,\textit{$^{a}\ddag$} Pablo Pe{\~n}as-L{\'o}pez,\textit{$^{b}\ddag$} {\'A}lvaro Moreno Soto,\textit{$^{c}\ddag$}} Devaraj van der Meer,$^c$ Detlef Lohse,$^c$ Han Gardeniers,$^a$ and David Fern{\'a}ndez Rivas $^{\ast a}$ \\
\vspace{1cm}

\normalsize{The formation, growth and detachment of gas bubbles on electrodes are omnipresent in electrolysis and other gas-producing chemical processes. To better understand their role in the mass transfer efficiency,  we perform experiments involving successive bubble nucleations from a predefined nucleation site which consists of a superhydrophobic pit on top of a micromachined pillar. The experiments on bubble nucleation at these spots permit the comparison of mass transfer phenomena connected to electrolytically generated H$_2$ bubbles with the better-understood evolution of CO$_2$ bubbles in pressure-controlled supersaturated solutions. In both cases, bubbles grow in a diffusion-dominated regime. For CO$_2$ bubbles, it is found that the growth rate coefficient of subsequent bubbles always decreases due to the effect of gas depletion. In contrast, during constant current electrolysis the bubble growth rates  are affected by the evolution of a boundary layer of dissolved H$_2$ gas near the flat electrode which competes with gas depletion. This competition results in three distinct regimes. Initially, the bubble growth slows down with each new bubble in the succession due to the dominant depletion of the newly-formed concentration boundary layer. In later stages, the growth rate increases due to a local increase of gas supersaturation caused by the continuous gas production and finally levels off to an approximate steady growth rate. The gas transport efficiency associated with the electrolytic bubble succession follows a similar trend in time. Finally, for both H$_2$ and CO$_2$ bubbles, detachment mostly occurs at smaller radii than theory predicts and at a surprisingly wide spread of sizes. A number of explanations are proposed, but the ultimate origin of the spreading of the results remains elusive.} \\
\vspace{12cm}

\makeatletter{\renewcommand*{\@makefnmark}{}
\footnotetext{\textit{$^{a}$ Mesoscale Chemical Systems Group, MESA+ Institute for Nanotechnology, Faculty of Science and Technology, University of Twente, P.O. Box 217, 7500 AE Enschede, The Netherlands}}\makeatother
\footnotetext{\textit{$^{b}$ Fluid Mechanics Group, Universidad Carlos III de Madrid, Avda. de la Universidad 30, 28911 Legan{\'e}s (Madrid), Spain}}\makeatother
\footnotetext{\textit{$^{c}$ Physics of Fluids Group and Max Planck Center Twente, MESA+ Institute and J. M. Burgers Centre for Fluid Dynamics, Faculty of Science and Technology, University of Twente, P. O. Box 217, 7500 AE Enschede, The Netherlands}}\makeatother

\footnotetext{\ddag These authors contributed equally to this work.}\makeatother
\footnotetext{$\ast$ Email: d.fernandezrivas@utwente.nl}\makeatother}
\footnotetext{\dag This article has been published in Energy \& Environmental Science: \url{http://dx.doi.org/10.1039/C8EE02657B}}

\pagebreak

\section{Introduction}
Hydrogen is a promising energy carrier that can be obtained via zero CO$_2$ emission techniques\cite{fujishima1972, barelli2008, chattanathan2012} such as solar-driven water splitting.\cite{maeda2005, maeda2007, jain2009, reece2011} However, the chemical reactions involved in such processes result in bubble generation. Such bubbles can block the reacting surfaces and decrease the process efficiency.\cite{zhang2012, dorfi2017}

The formation of bubbles on liquid-immersed surfaces is relevant for many gas-producing processes such as boiling,\cite{westwater1958} catalysis\cite{somorjai2010, chen2014} and electrolysis.\cite{dapkus1986, lv2017} More specifically, the formation of bubbles during chemical processes may be either beneficial due to increased heat and mass transfer induced by convection upon bubble detachment,\cite{zhang2011} or detrimental due to overpotentials caused by blocked active sites on the electrodes.\cite{sillen1983, dukovic1987, eigeldinger2000} 

\begin{figure*}
\centering
\includegraphics[width = 0.9\columnwidth]{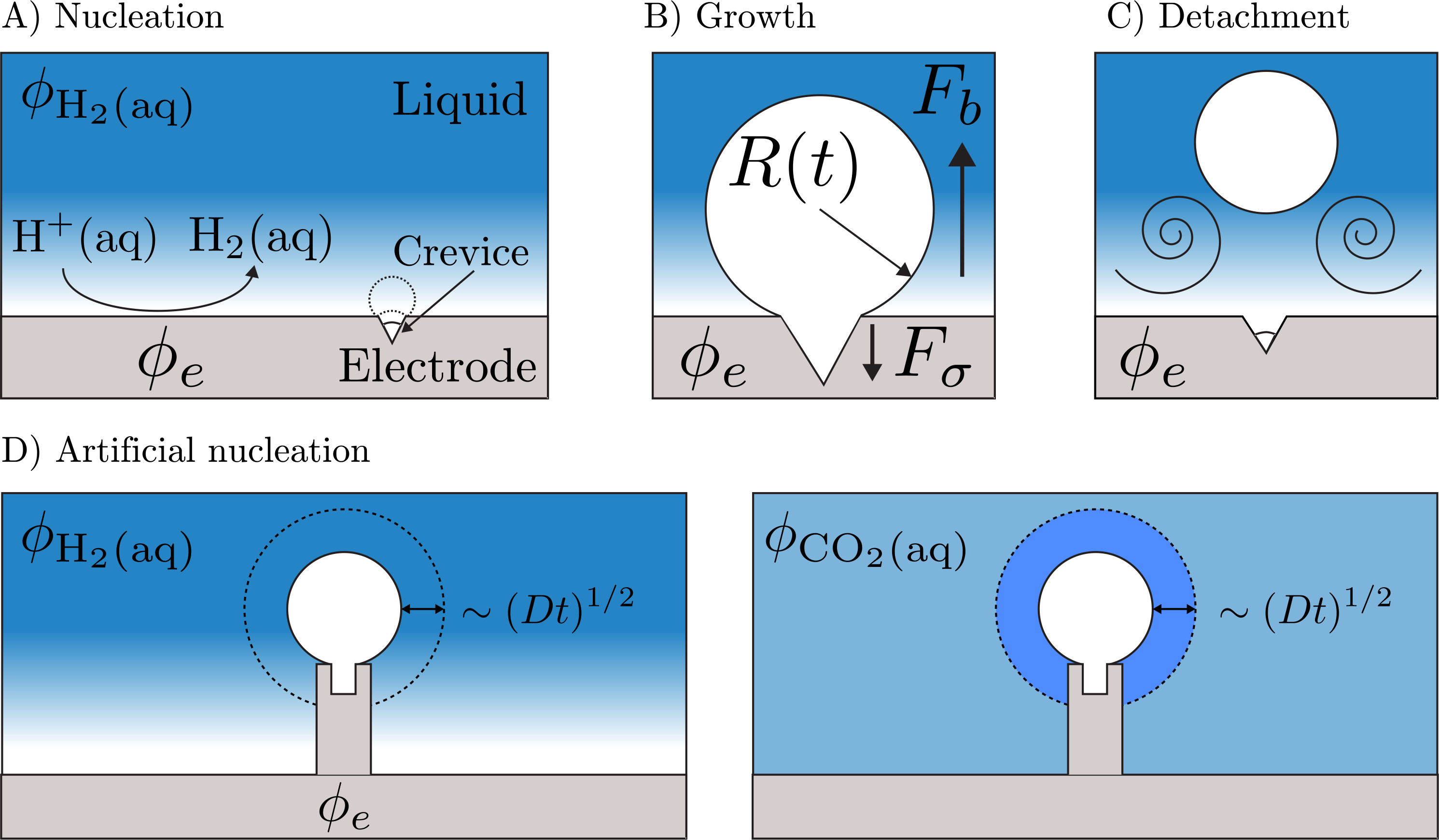}
\caption{Various stages of bubble evolution on electrodes. A) Heterogeneous bubble nucleation, here shown to occur in a crevice. The electron flux towards the electrode surface is indicated by $\phi_e$. The flux $\phi_\mathrm{H_2(aq)}$ indicates the diffusive transport of H$_2$ gas to the nucleating bubble. The highest gas concentration is at the electrode surface, indicated by a lighter blue color (the same colour pattern applies to the other plots). B) Bubble growth on the electrode surface. The direction of the interfacial tension force $F_\sigma$ and buoyancy force $F_b$ are shown. C) Detachment of bubbles by buoyancy overcoming the interfacial tension force which pins the bubble to a crack or crevice. D) Artificial nucleation sites to facilitate successive bubble evolution. On the left panel, the H$_2$ bubble evolution during water splitting is shown. The dotted area shows the time-dependent area from which the bubble experiences influx of gas via diffusion. On the right panel, the CO$_2$ bubble evolves in a CO$_2$ supersaturated medium. The gas concentration is homogeneous in the liquid apart from the time-dependent area around the bubble where the gas becomes depleted as successive bubbles grow,\cite{moreno2017} indicated by a darker blue color.}
\label{Figure1}
\end{figure*}

Bubbles preferably nucleate in small defects such as pits or crevices, where gas can be easily entrapped and the energy barrier is smallest.\cite{jones1999} A certain control over the location at which bubbles are prone to nucleate can be achieved by modifying the topography of the solid surface with suitable microstructures that act as preferential nucleation sites. The robustness of this concept has been demonstrated during pressure pulse propagation,\cite{bremond2006} ultrasound exposure,\cite{fernandez2010} turbulent boiling \cite{narezo2016} and under liquid flow conditions.\cite{gross2018} For this purpose, pillars are fabricated as preferential nucleation sites for bubbles, as shown in Figure \ref{Figure1}D, following a long-term line of research in our group with the aim of understanding and controlling the bubble evolution as a function of gas diffusion.\cite{enriquez2013, enriquez2014, moreno2017,vanderlinde2017}

Three different phases can be distinguished during bubble evolution as shown in Figure \ref{Figure1}: bubble nucleation at the surface (Figure \ref{Figure1}A), growth (Figure \ref{Figure1}B) and detachment (Figure \ref{Figure1}C).  In this study, we provide an in-depth comparative analysis between bubble evolution on a single pillar during electrolysis and the better-understood bubble evolution in pressure-controlled CO$_2$ supersaturated solutions on the same geometry, working out similarities and differences between the two processes. Our ultimate goal is to increase energy conversion efficiencies of solar-driven water splitting systems by controlling the gas bubble evolution on micromachined electrodes. 

\subsection{{\color{black}Outlook}} 
{\color{black}In this fundamental study, we have investigated the isolated bubble evolution on artificial nucleation sites micromachined on electrodes. The knowledge achieved with our experimental and theoretical work can certainly assist in the design of novel devices in the future. These future works could use nucleation sites to prevent the crossover of species in configurations in which the electrodes could be used to drive the bubbles to different streams \cite{hashemi2015} or to facilitate buoyancy driven separation mechanisms.\cite{davis2018} Artificial nucleation sites could also be used to evolve bubbles in predefined locations, a scenario which has been suggested to give rise to increased flexibility in device design, optimization and operation.\citep{ardo2018} The use of multiple nucleation sites on electrodes permits the definition of areas on the electrodes where bubbles are generated such that they do not compete for evolved gas as well as areas where they do. This could determine areas on the electrode surface where bubbles do not form, and dedicated areas where bubbles do form and would allow for controlled bubble formation at higher current densities. Major advantages could lie in designing electrodes where the catalytic surface is kept free from bubbles.}

\section{Materials and methods}

\subsection{Microfabrication of silicon substrates}
Micropillars on the surface of the electrode increase the active area and contact with the liquid phase, ultimate characteristics which are desirable in photolysis applications.\cite{elbersen2015, esposito2017} This approach encourages the construction of small and dense structures which work as light-harvesting areas. With the aim of understanding the fundamentals of bubble evolution on pillars, we focus on a single pillar microstructure of radius $R_p=2.5-15$ $\mu$m to study the succession of single bubbles generated on them. A superhydrophobic pit on top of the micropillar serves as the nucleation site.\cite{jones1999}

\begin{figure}
\centering
\includegraphics[width =0.65 \columnwidth]{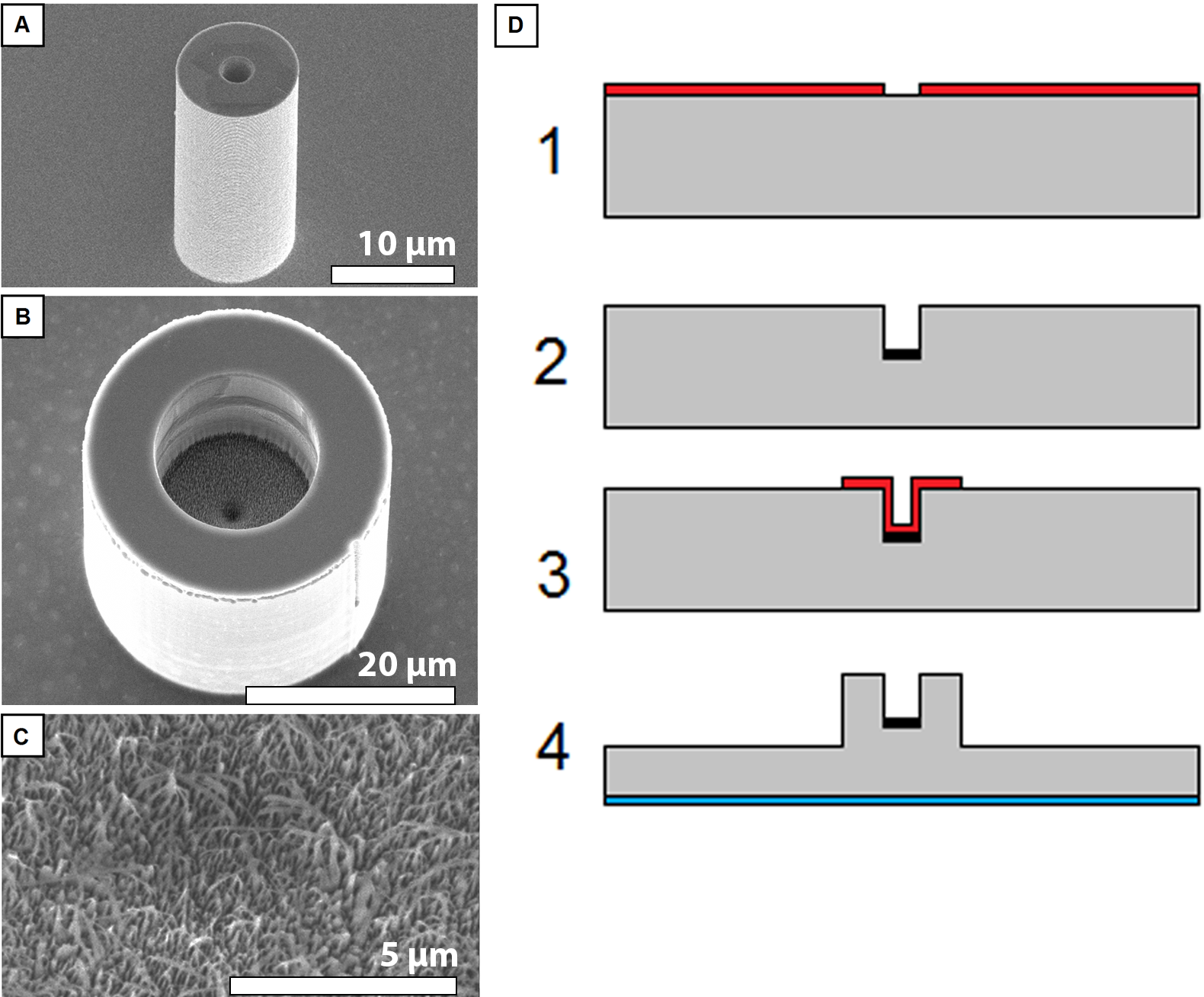}
\caption{Scanning Electron Microscope (SEM) images of A) a micropillar with a 10 $\mu$m diameter, a pit diameter of 2 $\mu$m and a pillar height of 25 $\mu$m, viewed at a 45$^\circ$ angle, B) a micropillar with a 30 $\mu$m diameter, a pit diameter of 15 $\mu$m and a pillar height of 30 $\mu$m viewed at a 20$^\circ$ angle, and C) a close-up of black silicon at the bottom of the pit in panel B viewed under a 20$^\circ$ angle. D) Sketch of the cross-sectional view (not to scale) of the substrate fabrication process. Step 1 shows the p$^{++}$ type silicon wafer on which a pattern is created via photolithography to mark the outline of the pit. With dry etching, a pit is created and black silicon formed at its bottom, step 2. Resist is applied and patterned via photolithography to mark the outline of the micropillar for dry etching, step 3, after which an aluminium backside contact is formed via DC sputtering.  The resulting complete substrate is shown in step 4.}
\label{Figure2}
\end{figure}

Boron-doped silicon wafers with (100) crystal orientation, resistivity in the range of 0.01 $\Omega\cdot$cm -- 0.025 $\Omega\cdot$cm, thickness of 525 $\mu$m and single side polished, were covered by 1.7 $\mu$m Olin OiR 907-17 resist. Using photolithography, circular regions ranging $R_0=1-10$ $\mu$m in radius were defined, as shown in step 1 in Figure \ref{Figure2}D. The circular regions were etched with a deep reactive ion etching (DRIE) Bosch process (Adixen AMS100SE) to a depth of $\sim$ 20 $\mu$m. Black silicon was formed at the bottom of the pits with DRIE, as shown in step 2 in Figure \ref{Figure2}D. Black silicon is an important structure that allows for better gas trapping while immersing the substrates in liquid. Afterwards, additional fluorocarbons were deposited ($\pm$ 40 nm/min) inside the pits, turning them superhydrophobic.\cite{borkent2009} The deposition times varied per set of samples between 7 s to 60 s.

The pillar radii were defined with photolithography as shown in step 3 in Figure \ref{Figure2}D. These pillars were etched with DRIE to various heights in the range of 0 $\mu$m -- 60 $\mu$m. An aluminium contact was created via DC-sputtering with a thickness of 100 nm (99$\%$ Al, 1$\%$ Si) at the bottom of the substrate, as shown in step 4 in Figure \ref{Figure2}D. An ultrasound (VWR Ultrasonic Cleaner USC-THD, 45 kHz) acetone bath was used to remove the resist. Afterwards, the wafers were diced (Disco DAD 321) into $10$ mm $\times10$ mm square substrates. Prior to the measurements, the samples were cleaned with another ultrasound acetone bath step. Figure \ref{Figure2}A-B shows SEM images of fabricated micropillars and Figure \ref{Figure2}C shows the black silicon inside the superhydrophobic pit. 

\subsection{Experimental set-ups for bubble evolution}
\begin{figure}
\centering
\includegraphics[width = 0.5\columnwidth]{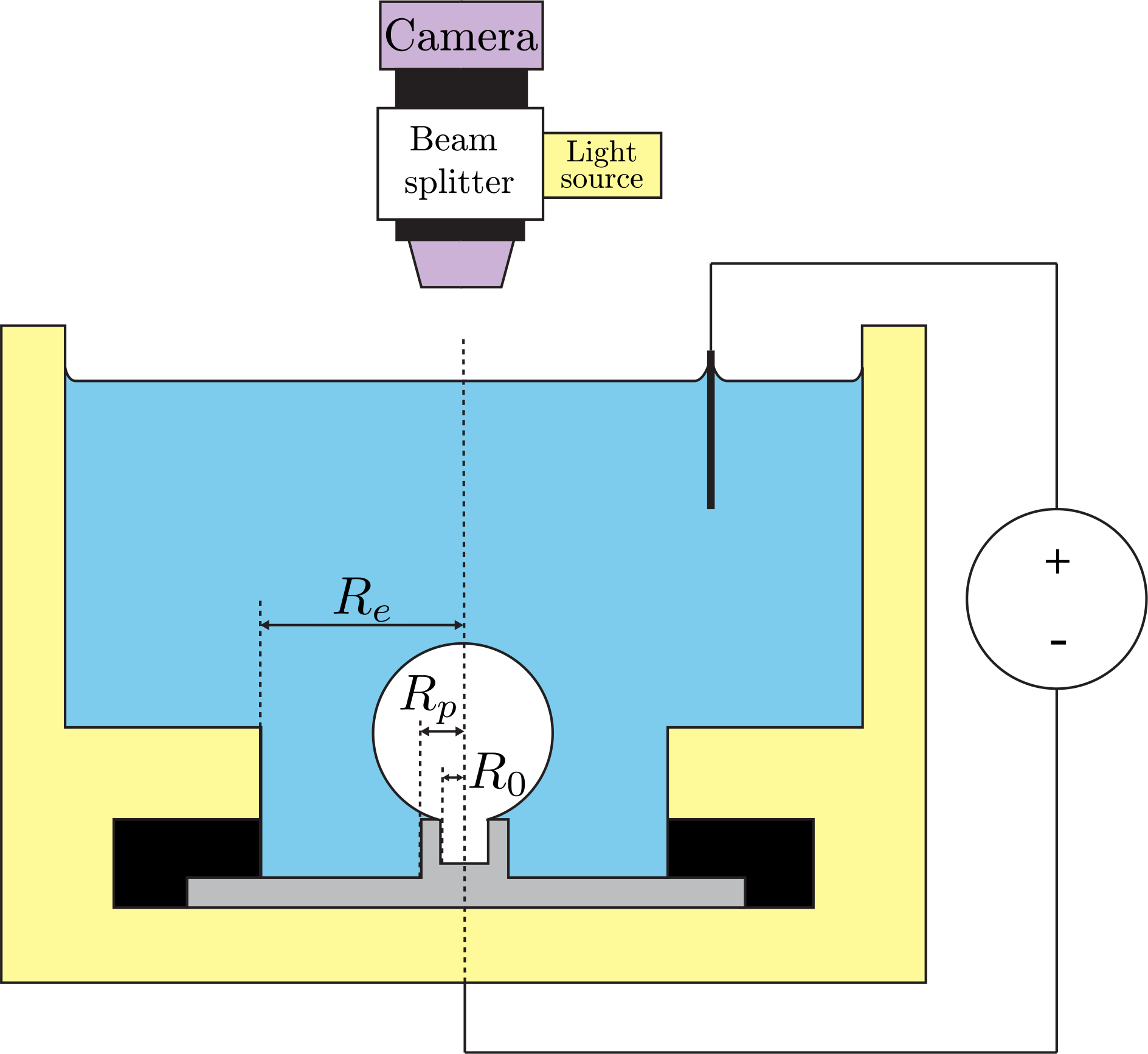}
\caption{Schematic of the electrolysis set-up (not to scale). At the top, the optics consist of a camera, lens and light source. Below the optics, an acrylic holder (yellow) which contains the substrate (grey) is placed. A circular area of the substrate of radius $R_e =$ 3.5 mm is in contact with the electrolyte (light blue), in which the counter electrode is placed (top right). A DC power source is used to drive the reaction.}
\label{Figure3}
\end{figure}

Figure \ref{Figure3} shows the electrolysis set-up, consisting of a custom-made acrylic holder, a camera and a power source. The acrylic holder is designed to keep the substrate in place, to hold a platinum wire counter electrode far away from the growing bubble and to contain the electrolyte. A circular area of the silicon substrate with radius $R_e = 3.5$ mm and sealed to the holder with a Teflon ring is in contact at all times with the electrolyte. This radius is approximately ten times the maximum bubble radius and, therefore, we can assume that the holder walls do not play any significant role during bubble growth on the pillars. The substrate contains an electrical contact at the bottom aluminium layer through which the current is supplied (not shown in Figure \ref{Figure3} for simplicity). A Keithley 2410 power source is used to drive the constant-current electrolysis. For optical imaging, a Flea\textsuperscript{\textregistered}3 Monochrome Camera, (optical resolution of 1.1 $\mu$m/pixel) is coupled to a 50/50 Beam-splitter Cube. For illumination, a Galvoptics KL2500 LCD 230V light source is used.

At the beginning of each experiment, the holder is filled with 20 mL of fresh electrolyte. The electrolyte consists of a solution of non-degassed Milli-Q water with 10 mM \ce{Na2SO4} salt and a pH 3 buffer of 1 mM anhydrous sodium acetate and 0.1 M acetic acid. The temperature remains constant at all times, $T\approx 20$ $^\circ$C. During each experiment, a constant current in the range of 10 $\mu$A -- 600 $\mu$A is supplied. The resulting current density $J$ falls in the range of 0.3  A/m$^2$ -- 15 A/m$^2$. {\color{black}The potentials during experiments were measured within a range of 1.8 V to 4.9 V.}

To compare the evolution of H$_2$ bubbles generated by electrolysis with that of CO$_2$ bubbles growing in an initially uniformly supersaturated solution, identical silicon substrates are placed within a pressurized test chamber (pressure $P_0 \approx 9$ bar) that is filled with carbonated water previously saturated at the same pressure. By lowering the pressure to approximate values of $P_l \approx 7.7$ bars, a supersaturation of $\zeta=P_0/P_l-1\approx 0.17$ is achieved following Henry's law (at constant temperature) and, consequently, bubbles nucleate and grow on the predefined spots. A detailed description of this experimental set-up and procedure can be found elsewhere.\cite{enriquez2013, moreno2017}

\section{Results and discussion}
\subsection{Bubble nucleation on a superhydrophobic pit}
The superhydrophobic pit entraps a gas pocket upon submersion in a liquid\cite{bremond2006} and hence acts as a site for heterogeneous nucleation. The interfacial or equilibrium concentration of dissolved gas at a liquid-gas interface can be written as $C=k_{H}P_g$ according to Henry's law, where $k_H$ is Henry's constant (a decreasing function of temperature specific to each gas-liquid pair) and $P_g$ is the partial pressure of the gas acting on the liquid surface.\cite{henry1803} For a pit with a circular opening of radius $R_0$, the pressure threshold at which bubbles begin to grow is given by the condition\cite{borkent2009}
\begin{equation}\label{eq:nucleation1}
P_v + P_g  > P_l +  \frac{2\sigma}{R_0} \equiv \frac{P_v}{P_l} +(\zeta+1) >  1 + \frac{2\sigma}{P_lR_0},
\end{equation}
where $P_v$ is the liquid vapor pressure, $P_l$ is the liquid pressure and $\sigma = 0.07$ N/m is the liquid-gas interfacial tension (for simplification, we assume a constant value for both H$_2$ and CO$_2$ cases). The radius $R_0$ in the Laplace pressure term (last term in (\ref{eq:nucleation1})) is justified since, at the nucleation stage, the bubble can be assumed to be a hemispherical cap of radius $R_0$ growing from the pit with the same radius. Equation (\ref{eq:nucleation1}) reflects that the pressure inside the bubble must overcome the forces resulting from the liquid pressure and surface tension to achieve bubble growth. If a multicomponent solution of $N$ volatile ideal gas species is considered, the condition for growth in (\ref{eq:nucleation1}) can be approximated as:\cite{tucker1975}
\begin{equation}\label{eq:nucleation2}
\frac{P_v}{P_l} + \sum_{i = 1}^{N}(\zeta_i +1) >  1 + \frac{2\sigma}{P_lR_0},
\end{equation}
where $\zeta_i = C_i/(k_{H,i}P_l) -1$ is the supersaturation of the dissolved gas species $i$ (in general, position and time dependent), with $C_i$ being the gas concentration in mol/m$^3$. With this equation, we can calculate the critical minimum supersaturation level required to overcome the energy barrier due to surface tension.

For the electrolysis case, we perform experiments at  $T  = 20$ $^\circ$C and $P_l = 1$ bar. Under these conditions, the water vapour pressure can be neglected since $P_v/P_l \sim 0.02$ (the effect of dissolved gases on the vapour pressure has been considered negligible since their mole fraction is small enough to assume that there is no appreciable change in the boiling point of water). H$_2$ gas bubbles grow in a binary solution of \ce{H2} and air since the electrolyte is not degassed {\color{black}(this condition is similar to that present in real electrolyzer applications)} and it is permanently exposed to ambient air throughout its preparation, subsequent storage and finally during experiments. Therefore, it is reasonable to assume that it is air equilibrated, i.e. $\zeta_\mathit{air} = 0$ (assuming that air is a single component entity). Consequently and according to (\ref{eq:nucleation2}), the minimum supersaturation of H$_2$ required to trigger growth for a typical pit radius $R_0=2.5$ $\mu$m corresponds to 
\begin{equation} \label{eq:zetah2}
\zeta_{\ce{H_2}} =  \frac{2\sigma}{P_lR_0} -1 \approx -0.44.
\end{equation}
In practical terms, the negative value above means that the presence of other dissolved species, i.e. air (which consists of a mixture of N$_2$, O$_2$ and other gases), makes bubble nucleation easier and, consequently, it is possible to achieve bubble nucleation shortly after initiating the electrolysis. We can anticipate that somewhat higher concentrations are required in practice. There are many other factors that can inhibit bubble nucleation and growth. Those will be discussed later in the text. 

In contrast, the experiments with CO$_2$ bubbles growing from pressure-controlled supersaturated carbonated water within a pressurized chamber are performed at a liquid pressure $P_l  \approx 7.7$ bars and isolated from the outside. The preparation procedure ensures that in the experimental chamber there are no other gas species present within the liquid apart from CO$_2$. Therefore, the minimum supersaturation required for nucleation is 
\begin{equation}
\zeta_{\ce{CO_2}} =  \frac{2\sigma}{P_lR_0}  \approx 0.07.
\end{equation}
Note that in this case a positive minimum supersaturation value is necessary. Supersaturation levels below $\zeta_{\ce{CO_2}} = 0.07$ were tried and resulted in no bubble generation. The lowest $\ce{CO_2}$ supersaturation for which we experimentally achieved bubble growth was indeed $\zeta_{\ce{CO_2}} \approx 0.07$.

\subsection{Bubble nucleation times} 
\begin{figure}
\centering
\includegraphics[width = 0.5\columnwidth]{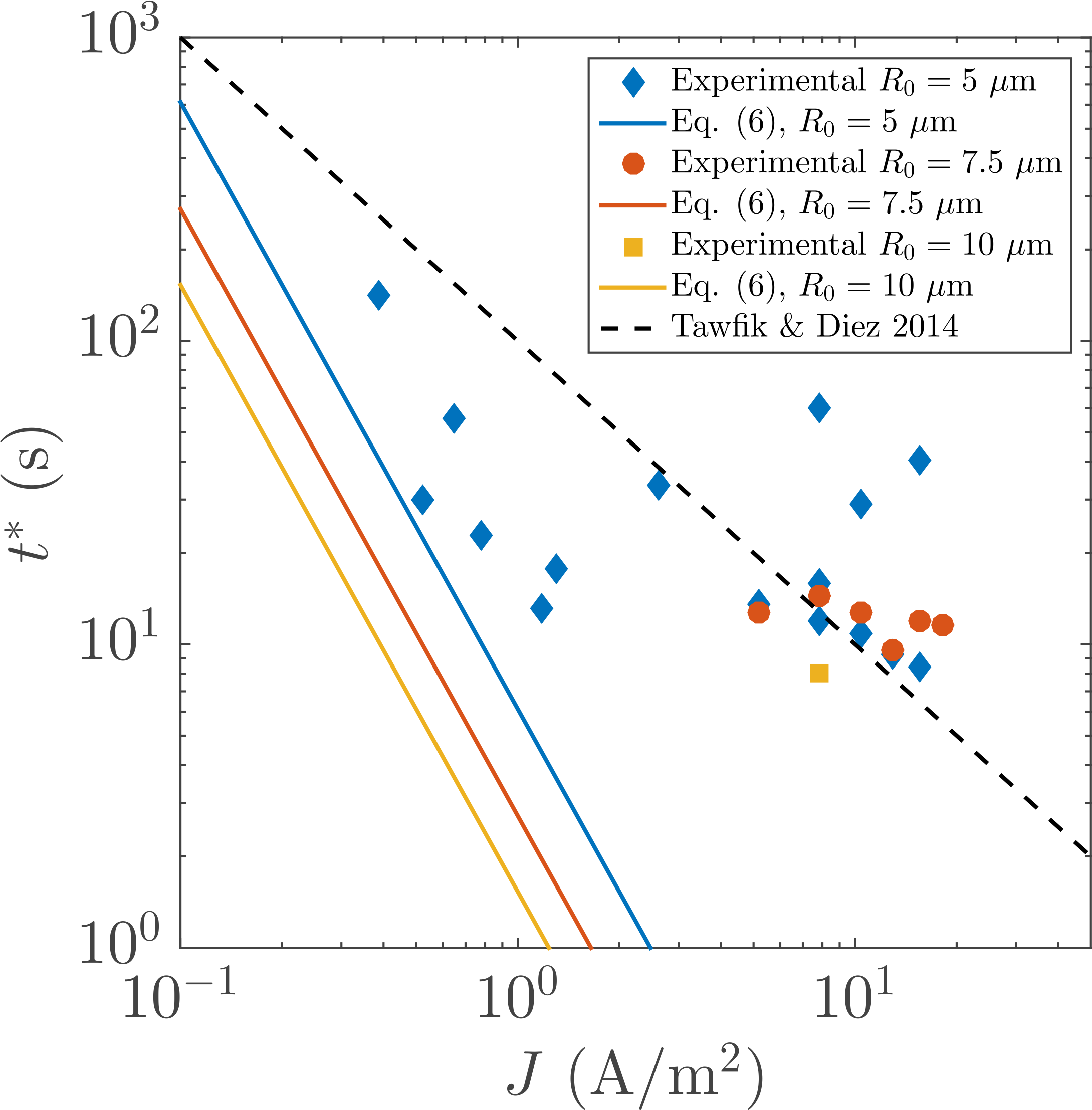}
\caption{Experimental nucleation time of the first H$_2$ bubble formed since the start of electrolysis as a function of current density. The blue diamonds, red circles and yellow squares show the measured nucleation times with $R_0=5, 7.5$ and $10$ $\mu$m, respectively. The blue, red, and yellow lines represent Equation (\ref{Eq:Nucleation_time_theo}) for $R_0=5, 7.5$ and $10$ $\mu$m, correspondingly. The dashed line shows the empirical relation by Tawfik and Diez (\ref{Eq:Nucleation_time}).\cite{tawfik2014} Generally, the nucleation time $t^*$ decreases with current density $J$. Discrepancies between experiments and the theoretical prediction are more than apparent and explained in the text.}
\label{Figure4}
\end{figure}

In constant-current electrolysis and in the absence of bubbles, the (molar) concentration of H$_2$ near the electrode can be estimated as  
\begin{equation} \label{eq:Celectrode}
C(t) = \frac{2J}{Fz\sqrt{\pi D}} \sqrt{t},
\end{equation}
which is an increasing function of time obtained by solving the 1D diffusion equation in a semi-infinite domain with a constant flux boundary condition.\cite{tawfik2014, vanderlinde2017} Here, $t$ denotes the time after the start of electrolysis, $J$ is the current density, $z = 2$ is the valency of the H$_2$ evolution reaction, $F = 96485$ C/mol is Faraday's constant and $D = 4.5\times10^{-9}$ m$^2$/s is the diffusivity of H$_2$ in water. Combining Henry's law, (\ref{eq:zetah2}) and (\ref{eq:Celectrode}), we obtain the theoretical minimum time for a bubble nucleation after the start of electrolysis as a function of the current density:
{\begin{equation}
t^* = \frac{\pi \sigma^2 k_{H,\mathrm{H}_2}^2 F^2 z^2 D}{ J^{2} R_0^{2}}.
\label{Eq:Nucleation_time_theo}
\end{equation}}
Here $k_{H,\mathrm{H}_2}$ $=$ $7.7 \times 10^{-6}$ mol/N$\cdot$m. It stands to reason that as $J$ increases, the gas formation rate also increases and, therefore, the minimum time to nucleate a bubble is achieved faster. There is evidence that the concentration at which the first bubble nucleates on a gas-evolving electrode also depends on the value of the current density.\cite{tawfik2014} \citet{tawfik2014} reported that the nucleation time does not depend on a constant concentration $C$, but rather on the applied current density $J$, with $C$ increasing as $J$ increases. They proposed the following empirical relation for the nucleation time of the first bubble spontaneously growing on a flat electrode in a presumably non-degassed electrolyte:
\begin{equation}
t^* = k \pi z^2 F^2 DJ^{-1}, 
\label{Eq:Nucleation_time}
\end{equation}
with $k = 0.19$ $\mathrm{mol^2/m^{4}A}$ a fitting constant. The nucleation times of the first H$_2$ bubble in the succession on the predefined pits are plotted vs the current density in Figure \ref{Figure4} and compared to the theoretical prediction in (\ref{Eq:Nucleation_time_theo}) with $R_0 = 5$, $7.5$ and $10$ $\mu$m, and the empirical relation in (\ref{Eq:Nucleation_time}). The times are measured from the start of the electrolysis up to a threshold radius of $\sim$25 $\mu$m, following the method used by Tawfik and Diez.\cite{tawfik2014} The nucleation time, $t^*$, appears to generally decrease with $J$; however, no clear trend can be appreciated. The significant variability in our experimental measurements can be attributed to three possible causes:
\begin{enumerate}
\item[(1)] The pit topography is different from sample to sample. We measured deviations from the ideal circular pit opening in the radial direction of several hundreds of nanometers (refer to the Electronic Supplementary Information). Fluorocarbons within the pit may hinder mass transport of dissolved gas towards the gas pocket or enforce pinning at different contact angles and, thus, affect the effective value of $R_0$.  
\item[(2)] The current density is likely far from being spatially uniform along the electrode surface.\cite{svetovoy2013, vanderlinde2017}
\item[(3)] {\color{black}The electrolyte contains air, partially composed of O$_2$. Oxygen reduction competes with H$_2$ formation. This implies that the net current density available for H$_2$ formation is less than the actually applied current density. By definition, the standard potential for H$_2$ formation is 0 V, whereas O$_2$ reduction occurs at $0.40$ V. Consequently, higher current densities result in both more H$_2$ production and O$_2$ reduction. This fact means that H$_2$ is not efficiently produced (not all of the applied current is used for its generation) and, thus, the bubble nucleation time seems not to follow a clear decreasing trend with increasing $J$. Furthermore, this may be a cause of the scattering in Figure \ref{Figure4}, since O$_2$ levels at the start of each experiment may not be the same (although a fresh solution was employed for each experiment). The levelling off of the nucleation time at higher current densities in the same figure could be attributed to the influence of the dissolved O$_2$ reduction, the unequal distribution of gas production, the time required for the diffusion of the gases through the liquid towards the artificial nucleation site and the stochastic nature of nucleation. In addition to the influence of the parameters mentioned above, other factors unknown to us may play a rather significant role in the measured deviation between the nucleation times of the bubbles and of the predicted theoretical values.}

\end{enumerate}
Moreover, the empirical prefactor in (\ref{Eq:Nucleation_time}) may correct for the growth of the bubble to the threshold size of 25 $\mu$m, even though the time needed to reach that threshold may be negligible compared to the time necessary to achieve nucleation. However, no such correction is performed in (\ref{Eq:Nucleation_time_theo}). Surface tension reduction due to dissolved gases in the solution can also explain why the experimental nucleation times differ from those predicted by theory.\cite{lubetkin2003} For electrolysis, the nucleation times for the various applied currents in this research fall within the order of tenths of seconds. In comparison, the nucleation of CO$_2$ bubbles in carbonated water is observed to occur at or below the order of seconds after the pressure was reduced below the saturation value.\cite{moreno2017} The differences may rely then on the different ways of bubble generation and not on the substrate surface properties.

\subsection{Bubble growth}
Bubble growth can be described as $R(t) \propto t^\alpha$, with $R$ denoting the bubble radius, $t$ the time after nucleation and $\alpha$ the time exponent.\cite{brandon1985} For diffusive bubble growth, $\alpha = 1/2$,\cite{epstein1950} whereas for reaction limited growth $\alpha = 1/3$.\cite{verhaart1980, brandon1985}

In electrolysis, diffusion-limited growth occurs when the characteristic time of the diffusive transport of the evolved gas across the electrode, $t_e\approx R_e^2/D$ (where $D$ is the diffusion coefficient), is much larger than that of the diffusive gas transport to the bubble, $t_d\approx R_d^2/D$.  The relation between this two characteristic diffusive times can be associated to the Damk\"{o}hler number, which is defined as $\mathit{Da} = t_e/t_d=R_e^2/ R_d^2$ and can be interpreted as the ratio of the diffusive transport across the characteristic electrode size and across the characteristic bubble size. Here, $R_e = 3.5$ mm is the electrode radius and $R_d \sim 0.3$ mm is the bubble experimental mean detachment radius of all experiments at all current densities, which results in $\mathit{Da} \approx 100$. Therefore, our research focuses on bubble growth during electrolysis controlled by diffusion. 

\begin{figure*}
\centering
\includegraphics[angle = 0, width=0.95\textwidth]{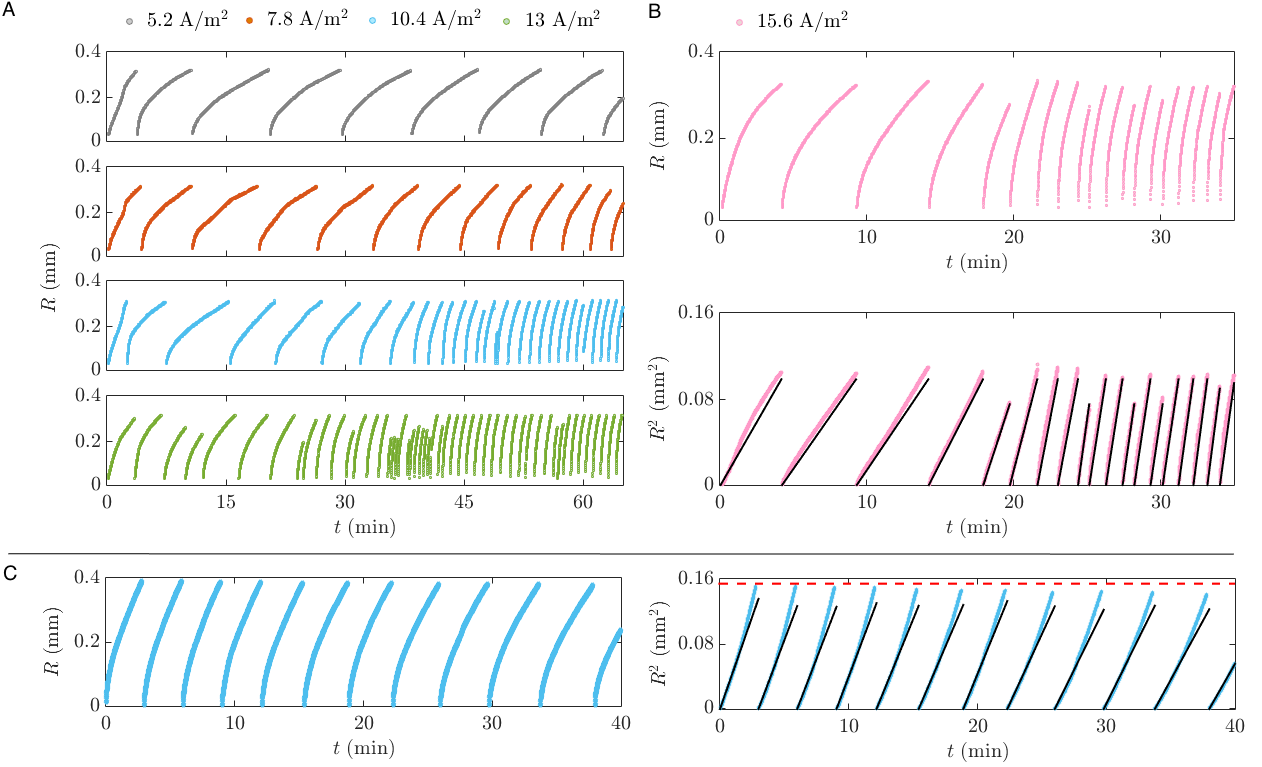}
\caption{\label{Figure5} A) H$_2$ bubble evolution on a microstructured electrode with a pit radius of $R_0 =5$ $\mu$m. The applied current densities are 5.2, 7.8, 10.4 and 13 A/m$^2$ from top to bottom, respectively. B) H$_2$ bubble evolution at 15.6 A/m$^2$ on a pit of radius $R_0 =5$ $\mu$m. The top figure shows the bubble radius as function of time whereas the bottom figure shows the experimental (pink) and theoretical (black) squared radii over time. C) Successive growth of CO$_2$ bubbles in supersaturated carbonated water on a pit of $R_0 = 10$ $\mu$m and supersaturation $\zeta$ $= 0.17$. The left plot shows the time evolution of the bubble radius. On the right plot, the experimental (blue) and theoretical (black) squared radii over time are shown. The dashed red line indicates the squared detachment radius of the first bubble. The onset of natural convection at the late stages of the bubble growth explains the deviation between the experimental and theoretical curves.\cite{enriquez2014}}
\end{figure*}

Figure \ref{Figure5}A and \ref{Figure5}B (top plot) show the evolution of bubble radii over time of five series of successive H$_2$ bubbles produced at constant-current electrolysis. Each series corresponds to a different current density. At the beginning, each successive bubble evolves slower than the previous one approximately up to the 4th bubble, when the growth rate becomes faster. This acceleration is attributed to the evolution of the diffusive concentration boundary layer in which the bubbles grow \cite{vanderlinde2017} and the most-probable complete reduction of the dissolved O$_2$ in the electrolyte (see item (3) in the discussion above). With increasing current densities, the growth rates at the beginning of each succession increase due to the larger gas production, but the evolution trend remains unaltered since the early bubbles in the succession deplete the diffusive concentration boundary layer around them. 

The unsteady nature of the electrolytic bubble growth becomes more apparent upon comparison with the bubble growth in pressure-controlled supersaturated carbonated liquid (Figure \ref{Figure5}C). In this figure, we present a succession of CO$_2$ bubbles in supersaturated water at $\zeta = 0.17$. The growth in this case continuously slows down with the successive bubble detachment due to the active depletion of the total amount of CO$_2$ gas available.\cite{moreno2017} In contrast, for the electrolytically-generated bubbles, after the early depletion the H$_2$ gas concentration near the substrate continues to increase over time due to the continuous water splitting reaction resulting in a faster growth of the H$_2$ bubbles.\cite{vanderlinde2017}

Both H$_2$ and CO$_2$ bubbles evolve via pure diffusive growth, namely 
\begin{equation}
R(t) = \tilde b\sqrt{Dt},
\end{equation}
where $\tilde b$ is the dimensionless growth coefficient.\cite{vanderlinde2017} The straight slopes observed in $R^2$ plotted against time in Figures \ref{Figure5}B (bottom plot) and C (right plot) corroborate this behaviour. {\color{black} A short movie showing a succession of single H$_2$ bubbles during electrolysis can be found on-line along this article (Movie 1).}

\begin{figure}[ht]
\centering
\includegraphics[angle = 0, width=0.5\columnwidth]{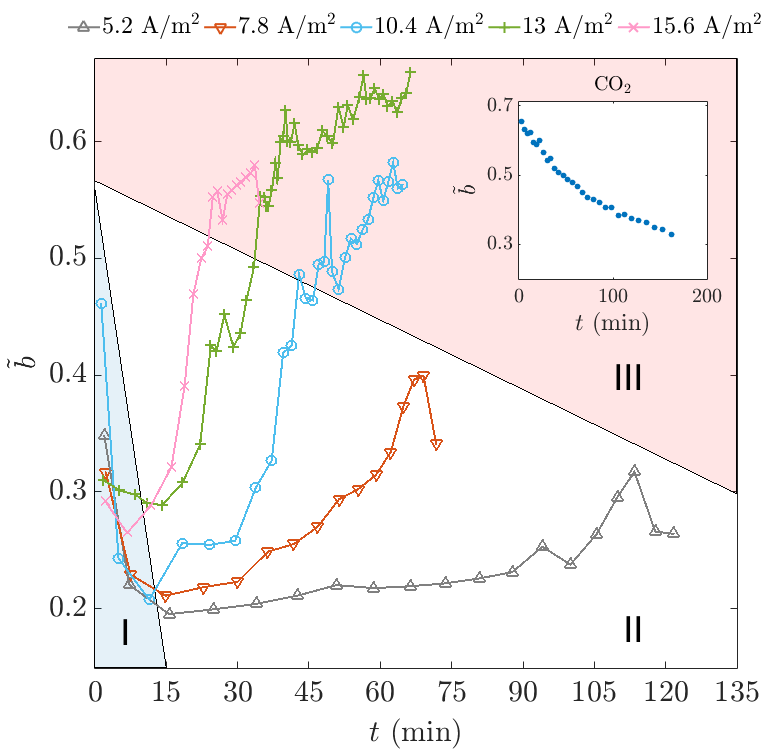}
\caption{\label{Figure6} The dimensionless growth coefficient $\tilde b$ per successive H$_2$ bubble as function of the time after the start of the electrolysis, $t$. The data are derived from the experimental results in Figure \ref{Figure5}A and the top plot in Figure \ref{Figure5}B. The different regimes are marked by the shaded regions, where region I corresponds to the stagnation regime in which $\tilde{b}$ decreases due to the early depletion, region II shows the counteracting effect due to the continuous gas production and III marks the regime in which an approximated steady state is reached. The transition between regimes II and III is defined by the moment in which the derivative $d\tilde{b}/dt$ drastically decreases, i.e. $\tilde{b}$ approaches a quasi-steady state. With increasing current density, the transition between regimes occurs faster because of the increased gas formation and the faster bubble evolution and their corresponding boundary layer. The inset shows the single regime I for the CO$_2$ bubble growth coefficient in supersaturated carbonated water caused by continuous depletion due to the successive bubble growth.}
\end{figure}

The gas boundary layer evolution during electrolysis results in three different growth regimes, which are further elucidated by taking a closer look at the growth coefficient $\tilde b$. Figure \ref{Figure6} shows the evolution of $\tilde b$ with time since the start of electrolysis calculated from the data in Figure \ref{Figure5}A and the top plot in \ref{Figure5}B. Note that each experimental point corresponds to the growth coefficient of a particular bubble in the succession. Initially, $\tilde b$ decreases as a consequence of the initial bubble locally depleting the boundary layer of gas, behaviour referred to as the `stagnation' regime (I). Successive bubbles keep growing in a mildly supersaturated liquid until the boundary layer overcomes the depletion losses due to the constant gas production and evolves to higher gas concentrations. The accompanying increase in $\tilde b$ characterises regime II, in which bubbles grow faster. The transition between regimes depends on the applied current density: the higher the current density, the earlier the onset of increasing $\tilde b$. Finally, regime III shows a stabilization in the growth rate for successive bubbles, reflected by $\tilde b$ increasing in small increments. In contrast, the growth coefficients corresponding to the CO$_2$ bubble succession in Figure \ref{Figure5}C always decrease due to gas depletion,\cite{moreno2017} inset in Figure \ref{Figure6}, similar to the early H$_2$ bubbles in electrolysis (regime I). In this case, there is no influx of new gas which can counteract this depletion effect, resulting in a continuous  smaller growth rate. The pillar height does not have any influence on the bubble growth coefficients.\cite{vanderlinde2017} For a more in-depth discussion on the different growth regimes and the influence that the boundary layer and its depletion have on the bubble growth dynamics, the interested reader is referred to \textit{van der Linde et al}.\cite{vanderlinde2017}

During both H$_2$ and CO$_2$ measurements, successive bubble growth could suddenly stop because of spurious pit deactivation. This may occur once liquid enters the pit during bubble detachment: the interface of the gas pocket in the pit can form a jet which can wet the surface inside the pit, displacing the air.\cite{borkent2009} We found no consistency in how long bubbles can be generated before pit deactivation. The fastest deactivation in the measurements occurred after the growth and detachment of a single bubble.

\subsection{Bubble detachment}\label{Sec:cl}
\begin{figure*}[ht]
\begin{centering}
\includegraphics[width=0.6\textwidth]{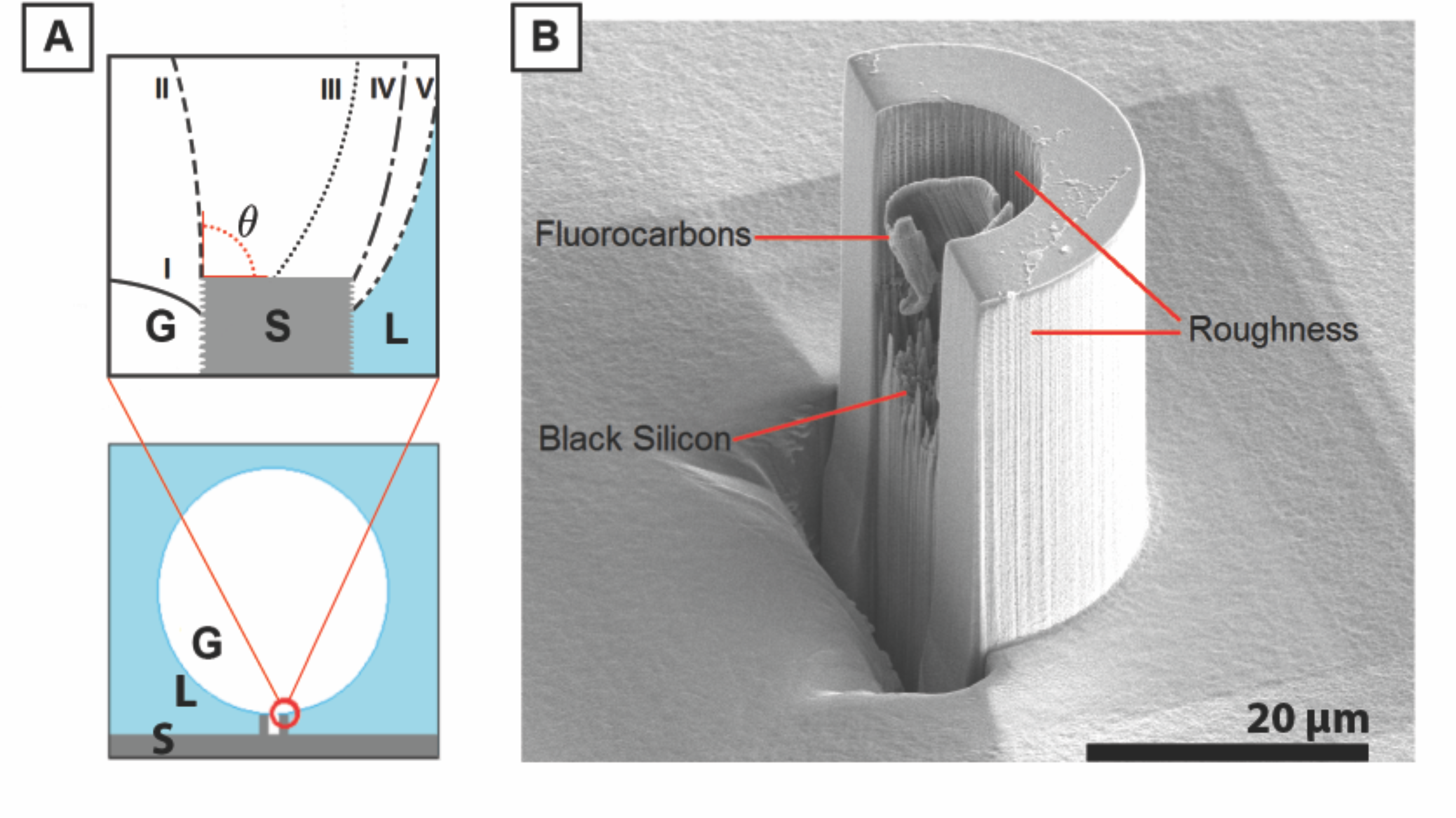}
\caption{A) Schematic side view of a cross section showing a micropillar with micropit, indicating five possible pinning positions for the contact line of evolving bubbles. Position I is inside the pit, II is at its edge with an arbitrary contact angle $\theta$ shown in red, position III indicates a transition location, IV the outer edge of the pillar and V is on the pillar surface. B) Focused ion beam (FIB) milled down pillar, under 52$^\circ$ angle with respect to the electron source. The inside of the pit shows the black silicon needle structure at the base and a detached fluorocarbon layer. Etching defects (vertical lines along the pillar) on the inside and on the outside are present. A video recording of the milling process can be found on-line along this article {\color{black}(Movie 2)}.}
\label{Figure7}
\end{centering}
\end{figure*}

The position of the triple contact line on the pit-pillar microstructure and the contact angle dynamics determine the size at which the bubbles detach from the microstructure.\cite{dettre1965, degennes1985} Since optical access to the contact line was not possible, we speculate on five probable pinning positions during the bubble evolution process, sketched in Figure \ref{Figure7}A. The inner surface of the pit contains several artifacts as a result of the fabrication process that can pin the bubble interface. As shown in Figure \ref{Figure7}B,  needle-like structures of black silicon are present at the bottom of the pit, whereas the inner surface contains vertical and horizontal scallops resulting from the Bosch etching process.\cite{park2003} Additionally, the fluorocarbon (FC) layer deposited for enhanced hydrophobicity can facilitate pinning. Typically, the FC layer will adhere to the pit wall; however, in Figure \ref{Figure7}B the layer detached prior to the FIB milling process (a video can be found on-line along this article) as observed with optical microscopy and SEM. This event could provide unpredictable pinning positions during the experiments and, consequently, end up in a different detachment radius. However, we have evidence that for the majority of the bubbles, the pinning is most likely to occur inside the pit (position I) throughout their whole lifetime, forming a bridging neck between the gas trapped in the pit and the bubble growing outside.\cite{moreno2017}

As the bubble grows and attains its detachment size, it is possible that the bubble contact line moves from position I up to V,\cite{ramos2006} as sketched in Figure \ref{Figure7}A. The departure size is an indirect way of estimating the position of the contact line. The maximum theoretical value of the bubble detachment radius growing from a pit of radius $R_0$ is given by Fritz's formula,\cite{fritz1935}
\begin{equation}
R_d^* = \left( \frac{3R_0\sigma}{2\Delta \rho g}\right)^{1/3},
\label{eq:Fritz}
\end{equation}
with $\Delta\rho$ the difference in density between the liquid and gas phases and $g = 9.81$ m/s$^2$ the gravitational acceleration. Equation (\ref{eq:Fritz}) can be derived from the balance between buoyancy and capillary forces, assuming that the contact line is at position II with a contact angle of 90$^\circ$ with respect to the horizontal at the moment of detachment, as sketched in Figures \ref{Figure1}B and \ref{Figure7}A. Net charges present on bubbles due to the solvent pH or absorbed species, such as surfactants, \cite{brandon1985b} may affect the pinning position of the bubble to the pit and consequently, its final detachment radius. Our electrolysis experiments are carried out in a medium with a pH 3 buffer and with no absorbent species to ensure a point of zero charge on the bubble. We can thus exclude electrostatic forces from the detachment force balance. 

\begin{figure} 
\centering
\includegraphics[angle = 0, width=0.55\columnwidth]{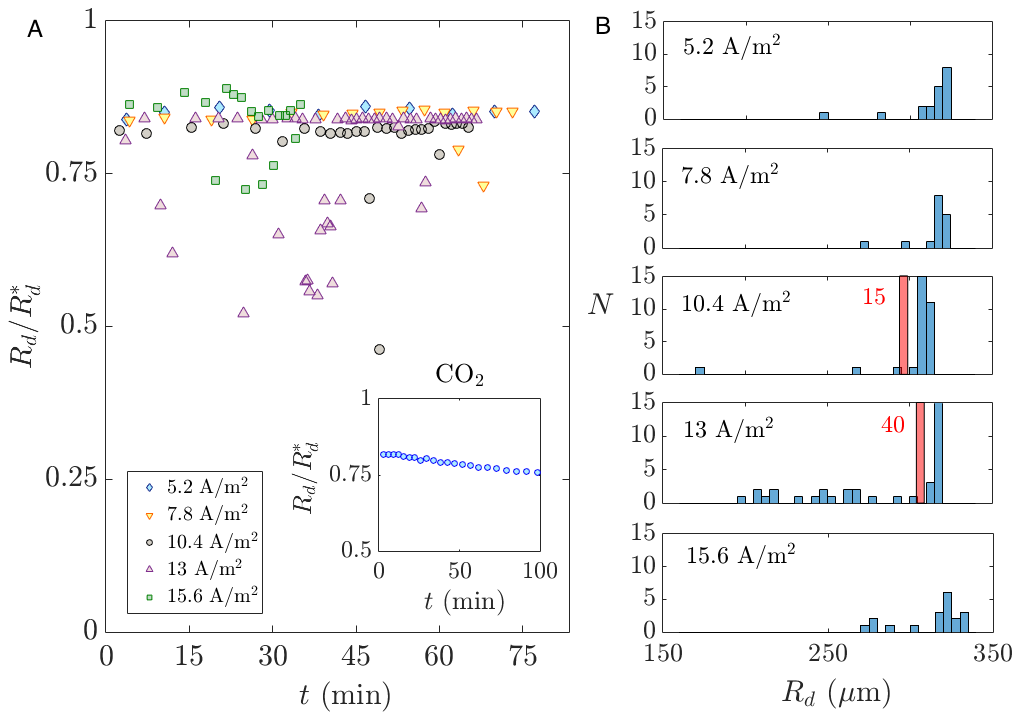}
\caption{\label{Figure8}A) Successive H$_2$ bubble detachment radii over time, normalized by the Fritz radius (\ref{eq:Fritz}) for various levels of applied current density. The inset shows the normalized detachment radii of successive CO$_2$ bubbles over time for $\zeta = 0.17$. In this case, the detachment radii slightly decrease with time due to induced density-driven convection, which does not occur in electrolysis. B) Histograms of the measured detachment radius for various applied current densities (from top to bottom, 5.2 A/m$^2$, 7.8 A/m$^2$, 10.4 A/m$^2$, 13 A/m$^2$ and 15.6 A/m$^2$). The red bars at current densities 10.4 A/m$^2$ and 13 A/m$^2$ have values of $N= $15 and $N= $40, respectively.}
\end{figure}

Figure \ref{Figure8} shows the detachment radius for electrolysis at various current densities. The measured radii are smaller than what equation (\ref{eq:Fritz}) predicts, as one would expect from the contact line pinned somewhere inside the pit (position I) and a potential necking process.\cite{moreno2017} Histograms of the detachment radii per current density applied to the same sample are included in Figure \ref{Figure8}B. The detachment radius does not seem to be affected by the current density. The inset in Figure \ref{Figure8}A likewise shows that the detachment radii of successive CO$_2$ bubbles always fall below the theoretical value. Moreover, the measured radii slightly decrease with each successive bubble formed due to the onset of buoyancy-driven convection near the bubbles.\cite{enriquez2014}  However, in general terms, bubble detachment radii remain stable and reproducible, especially in the short term (below $1$ hour).\cite{moreno2017} 

\begin{figure}[ht]
\centering
\includegraphics[angle = 0, width=0.5\columnwidth]{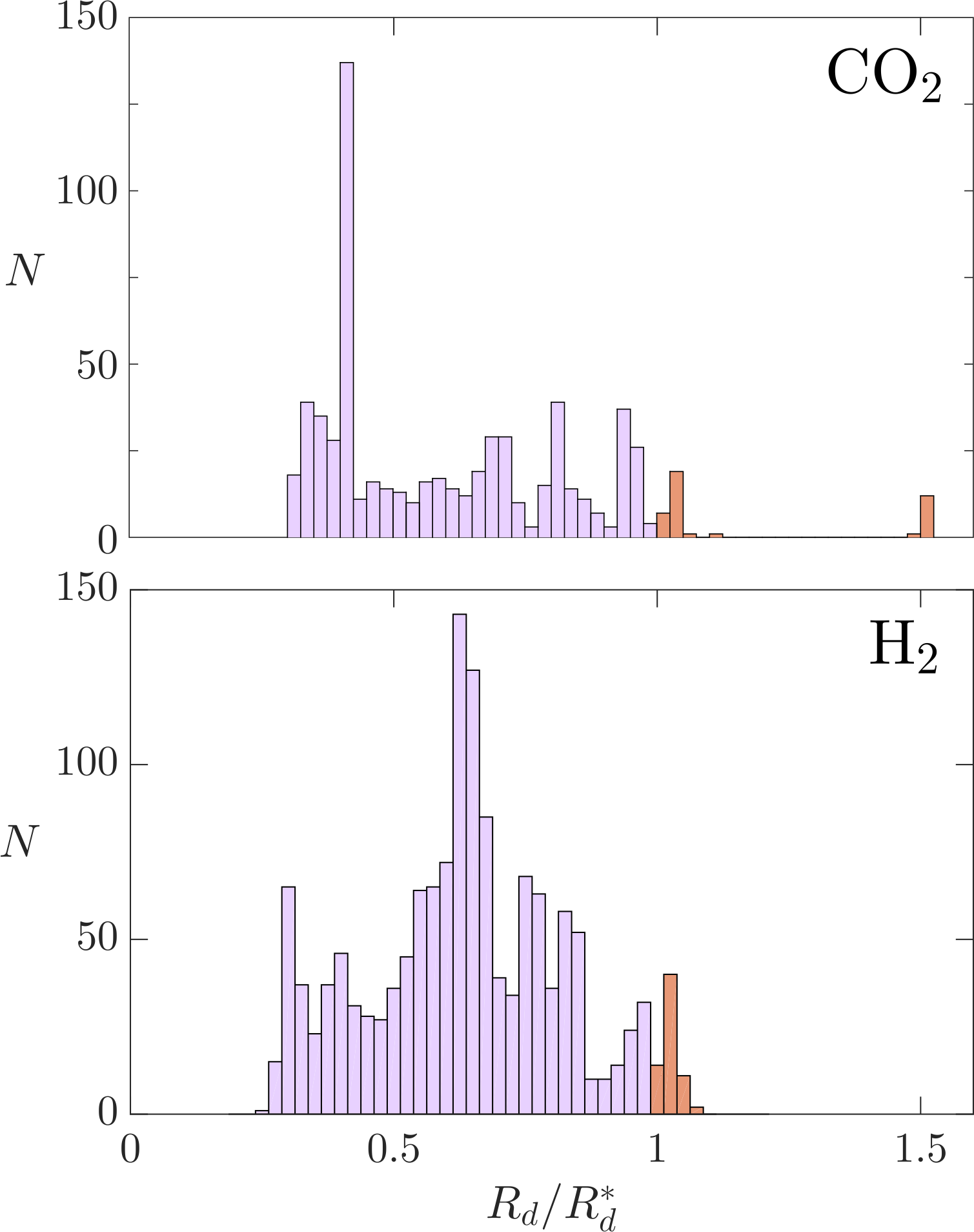}
\caption{\label{Figure9} The top histogram shows the detachment radii of CO$_2$ bubbles normalized by the Fritz radius (\ref{eq:Fritz}) formed at 0.16 $ < \zeta <$ 0.18. Values below $R_d/R^*_d$ $<$ $1$ are shown in purple whereas larger values are shown in red. The bottom histogram shows the detachment radii of H$_2$ bubbles evolved on various substrates at different current densities. Here the same color palette as in the top figure has been used. Even though the histograms have different distributions, both correspond to a mean value of bubble detachment $R_d/R^*_d \approx 0.6$.} 
\end{figure}

Figure \ref{Figure9} shows two histograms of the detachment radii $R_d$ normalized with the theoretical maximum detachment radius $R^*_d$. The top histogram shows that the detachment radii of CO$_2$ bubbles spread over a range with a mean value of $\sim 0.6$ $R_d/R^*_d$. The lower histogram shows a peaked distribution for the H$_2$ bubble detachment radii, also with a mean value of $\sim 0.6$ $R_d/R^*_d$. The same mean range of $R_d/R_d^*$ for the H$_2$ and CO$_2$ measurements is not accidental since both scenarios make use of similar microstructures with the same pit-pillar configuration. The spread in the measured radii must arise from the fact that the contact line may differ from experiment to experiment, and thus the necking before pinch-off occurs differently. Preferred adhesion sites or defects within the pit or on the pillar could be responsible for this. Since roughness of flat electrodes has been shown to influence the detachment radii of bubbles,\cite{baum1998, barker2002, krasowska2007} we expect that pit roughness might play a role in the detachment radii of evolving bubbles. We measured the roughness in radial direction but found no apparent correlation between the detachment radii and the radial roughness (see the Electronic Supplementary Information). For some bubbles, $R_d/R^*_d > 1$, probably due to the fact that the bubbles were not pinned to the pit (positions I or II in Figure \ref{Figure7}A) but rather to defects on the pillar or the outer rim (position III, IV or V in Figure \ref{Figure7}A). In our experiments, we have measured detachment radii up to $1.5R^*_d$, especially for the case of the smallest pit to pillar radii ratio. This case is particularly interesting, since such a small ratio could be used for future designs of pillars in which the pit functions as the gas trapping source and the pillar as the outer pinning geometry for the bubble. Convective forces, electrostatic charges induced by local pH changes, and the dependency of surface tension and liquid density with concentration of dissolved gases may also influence the force balance and final detachment radius in a complex way. {\color{black} Although we provide several possible scenarios and parameters which could cause the deviation between the measured detachment radii and theory, the influence of other unknown factors can not be excluded.} Nonetheless, a full analysis of the force balance and other factors influencing detachment is beyond the scope of this study.

\subsection{Gas transport efficiency}\label{Sec:Eff}
The efficiency of gas transported away from the electrode surface by the bubbles can be quantified as the ratio between the amount of gas moles within each bubble after detaching from the nucleation site, $n_b$, and the total amount of electrolytically produced moles of H$_2$, $n_g$. Note that this efficiency is not constant in time since it changes as the subsequent bubbles grow at different rates and depends on the amount of dissolved O$_2$ which is reduced at the electrode. The efficiency after the $n$-th bubble in the succession has detached is thus calculated as
\begin{equation}
\displaystyle\frac{n_b}{n_g} = \frac{\mathlarger{\mathlarger{\sum}}\limits_{i=1}^n{ 
\displaystyle\frac{4 \pi P_{b,i}R_{d,i}^3}{3R_u T } } } {Q/(Fz)}.
\label{Eq:Faraday}
 \end{equation}
Here, $R_{d,i}$ denotes the detachment radius of the $i$-th bubble, $P_{b,i} = (2\sigma /R_{d,i}) + P_l$ is the internal pressure of the $i$-th bubble, F $= 96485$ C/mol, $z= 2$ is the valency of the H$_2$ evolution reaction, $Q(t_n) = J \pi R_e^2 t_n$ is the total electric charge supplied at the detachment time of the $n$-th bubble $t_n$, $R_u = 8.314$ J/K$\cdot$mol is the universal gas constant and $T = 293$ K the absolute temperature. Note that this definition of the efficiency is limited to the gas transported away in each bubble and, therefore, does not consider the gas transport from the electrode in the form of convective plumes caused by bubble detachment or in the form of parasitic bubbles growing in other spots within the set-up.

\begin{figure}
\centering
\includegraphics[angle = 0, width=0.5\columnwidth]{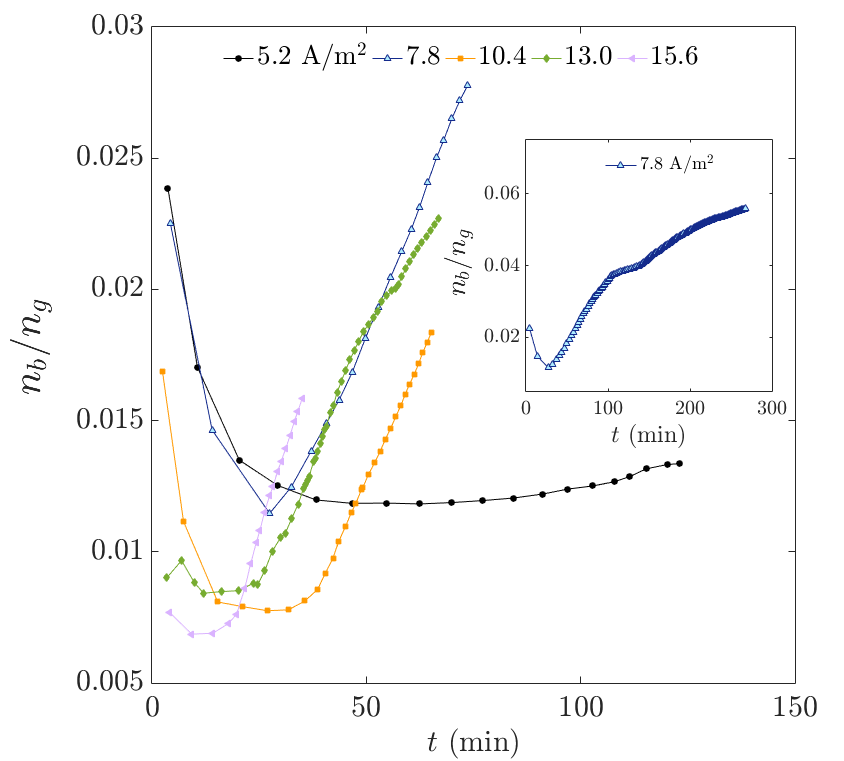}
\caption{\label{Figure10} The ratio of gas transported out of the liquid phase by the bubbles and the amount of electrolytically generated gas as function of time for various current densities. For the measurement at 5.2 A/m$^2$, the current density is so low that only regime I appears within our experimental time. It is expected that the other regimes (II \& III) would occur with prolonged reaction time. The inset shows the efficiency ratio for the full length of the 7.8 A/m$^2$ measurement up to 270 minutes. A maximum efficiency of 5.7 \% is obtained at the end of the experiment. The employed nucleation site has a radius of $R_0$ = 5 $\mu$m.}
\end{figure}

Figure \ref{Figure10} shows the H$_2$ transport efficiency of the bubble succession as a function of time. A single substrate is used for the measurements of the various current densities. The efficiency evolves as a parabola in time for all current densities, i.e. a similar trend as that of $\tilde{b}$ in time, Figure \ref{Figure6}. This originates from the definition of the transport efficiency, equation (\ref{Eq:Faraday}), which fundamentally corresponds to a discrete integral of $\tilde{b}$ in time. Consequently, the efficiency initially decreases due to the effect of depletion during the stagnation regime, region I in Figure \ref{Figure6}. During the stagnation, the efficiency is surprisingly higher for lower current densities. This may originate from larger depletion losses caused, for instance, by the formation of parasitic bubbles. However, the efficiency becomes larger with increasing current densities as the concentration boundary layer evolves with time to higher gas concentrations. This is expected since the current density is directly proportional to the generation rate of molecular hydrogen. The produced gas does not diffuse fast enough into the bulk electrolyte, but accumulates instead around the bubble and electrode, increasing the local supersaturation. This results in faster bubble formation frequencies and higher transport rates. We find the highest experimental efficiency (5.7 \%) for a current density of 7.8 A/m$^2$ after 270 minutes of constant electrolysis operation. A general optimal efficiency value could not be determined due to the eventual pit deactivation or parasitic bubble formation blocking optical access.

Future designs of electrodes with multiple nucleation sites may increase the amount of gas that is transported away by the bubbles, resulting in higher transport efficiencies. The size of the nucleation sites and the spacing over the surface would be crucial since they determine to what degree the bubbles compete for gas and how the gas concentration boundary layer evolves with time.

\section{Conclusions}
The microfabrication of artificial nucleation sites (in the form of pillar-pit microstructures on flat silicon substrates) allowed us to experimentally study bubbles evolving in water. By observing the succession of single bubbles, we compared the differences between the pressure-controlled supersaturated CO$_{2}$ and electrolytic H$_{2}$ bubbles, focusing on the evolution of the concentration boundary layer and its effect on the bubble growth rate, the detachment radius and the gas transport efficiency. 

The time taken for the first H$_2$ bubble to nucleate after the start of electrolysis at various current densities coincides with previous electrolysis nucleation studies and covers a wide spread ranging from the order of seconds to tens of seconds (most probably affected by the presence of dissolved O$_2$ at the beginning of the experiment) whereas the CO$_2$ nucleation occurs generally in the order of seconds once the carbonated solution becomes supersaturated. By studying the growth coefficient $\tilde{b}$, we determine that a system with a finite amount of gas available will experience continuously slower bubble evolution over time due to gas depletion, whereas in the case of electrolytically generated bubbles, their growth experiences different phases depending on the concentration of available gas as a function of time. The height of the pillars does not seem to play any significant role during bubble evolution in any of the cases studied here.

Bubble detachment usually occurs around 60\% of the maximum theoretical radius (see equation (\ref{eq:Fritz})) for both cases. This fact indicates that bubble detachment is mainly governed by the pillar-pit geometry. The smaller detachment value originates from the structural imperfections of the pits that lead to random adhesion sites of the contact line. The contact angle, the force balance and the neck formation of the bubbles are thus affected. For CO$_2$ bubbles, detachment occurs at slightly decreasing radii over time because of the onset of density driven convection \cite{enriquez2014} and a neck formation between the trapped gas in the pit and the growing bubble on top.\cite{moreno2017} In electrolysis, the detachment of H$_2$ bubbles does not follow any clear trend.

Finally, the gas evolution efficiency follows a parabolic trend with time. A matching trend is observed for the bubble growth rates. We conclude that the efficiency first decreases due to depletion losses, and then increases after a certain supersaturation is achieved and the dissolved O$_2$ is reduced. Surprisingly, during the stagnation regime the efficiency is higher for lower current densities. This effect is counteracted later in time, such that higher current densities $J$ imply higher efficiencies. The maximum efficiencies range from 1 to 5 \%, values which could be further increased with the use of multiple nucleation sites and flow conditions, closer to real life applications where continuous flow reactors are desirable. The aspects of nucleation, growth, and detachment considered here certainly warrant future studies toward higher transport efficiencies of (photo)electrolytic devices.

\section*{Conflicts of interest}
There are no conflicts to declare.

\section*{Acknowledgements}
We would like to thank S. Schlautmann for the discussion and fabrication of the experimental substrates, and R. P. G. Sanders for the discussions regarding the electrolysis set-up. We would further like to extend our thanks to H. A. G. M van Wolferen for the FIB milling and SEM imaging and the MESA+ Nanolab for the use of their facilities. This work was supported by the Netherlands Center for Multiscale Catalytic Energy Conversion (MCEC), an NWO Gravitation programme funded by the Ministry of Education, Culture and Science of the government of the Netherlands.

\balance

\bibliography{rsc} 
\bibliographystyle{rsc} 

\end{document}